 \documentstyle[sprocl]{article}
 
  \bibliography{unsrt}
\begin{document}
 \title{FRACTAL LOCAL GALAXY NUMBER COUNTS MAY IMPLY STRONG BIAS}
 \author{D. R. MATRAVERS}
 \address{School of Computer Science and Mathematics, University of 
 Portsmouth, Portsmouth PO1 2EG, UK}
 \maketitle
 \abstract
Exact Tolman solutions are used to analyse the implications if the
galactic number count  has a fractal form out to a distance of about 
150 Mpc in a universe which is homogeneous on the large scale.  It is
concluded that such a model requires either a non-linear Hubble law
or a very low density if galaxies trace the total matter
distribution.

 \section{Introduction}

 \noindent This paper is based on joint work with Roy Maartens and 
Neil 
 Humphreys.
 More details are given in three joint papers \cite{RM} \cite{NP} \
 cite{DR}.  There is
 substantial evidence \cite{PE} \cite{PI} to support a claim that the 
 number of galaxies
 within a volume defined by a distance $y$ has a fractal distribution,
 \begin{equation}
 N(y) = Ay^{\nu}   \label{fra}
 \end{equation}
 where $0 < \nu < 3$ is the fractal index and $A$ is a constant.  How 
 large $y$ can be is
 a matter of dispute but there is some agreement that for $y < $ 150 
Mpc 
 the formula
 (\ref{fra}) holds with $\nu$ between 1.5 and 2.

Unfortunately at this stage in the history of cosmology there is no
convincing theoretical model which predicts such a distribution.
Some work has been done recently \cite{VE} but it is only a
beginning.  This is a problem because it is notoriously difficult
to derive a convincing theoretical model from data alone.
Statistical procedures are excellent if we know the model and wish
to determine the parameters.  However they become suspect when we
have to derive the model and the parameters from the same data.
Accumulation of data and improved statistics \cite{PI} help but
they do not remove unease and doubt, e.g. is the distribution
really multi-fractal or is the data set incomplete?  We can expect
the debate to continue.

Our contribution has been to provide a different approach to the
problem.  We assume; (a) a fractal form for the number counts holds
to a distance (observer area distance) 150 Mpc from us, (b)
space-time is spherically symmetric about us, (c) Einstein's field
equations hold, (d) the matter distribution can be modelled as dust
on a suitable averaging scale, (e) beyond 150 Mpc the universe can
be modelled by a Friedmann-Lem\'{a}itre-Robertson-Walker (FLRW)
geometry, (f) the solutions are regular, and there is no shell
crossing or surface density layers and (g) the Darmois conditions
hold at matching surfaces.  A consequence of these assumptions is
that we can use a Tolman model within 150 Mpc.  Consistent with the
picture we take seriously that the observations are on the
observer's past light cone. This is often not done by astronomers
as pointed out by Ellis et al \cite{ST} and Laix and Starkman
\cite{DE} and theorists do not always do it either \cite{RA}, but
it has significant effects, for instance on the measurements of the
density power spectrum as pointed out by Laix and Starkman
\cite{DE}.  Even in the spatially homogeneous FLRW models it has a
non trivial effect leading to an in-homogeneous number count
formula \cite{EL} \cite{RI} \cite{ST}.

Observational coordinates \cite{ST} provide a powerful tool for
analysis on the past light cone of an observer and hence for
discussions of real observations \cite{MA} \cite{QJ}. They are
especially well suited
to spherically symmetric space-times.  In these coordinates the
null geodesic generators of the observer's past light cone are
trivially integrated whereas they cannot be integrated exactly for
Tolman models in $(3 + 1)$ coordinates \cite{BO} \cite{BN}
\cite{TO}. The reverse situation holds for the field equations
\cite{RM}. It follows that, in general,  the transformation between
observational and $(3 + 1)$ coordinates cannot be integrated
exactly in simple functions.  Thus, as could be anticipated, the
two coordinate systems give different viewpoints on the geometry
and the physics; the $(3 + 1)$ coordinate approach provides the
physical interpretations and the observational coordinates relate
to astronomical observations.  In this work we needed both
viewpoints and therefore found it necessary to develop results
which enable us to flip between the two systems
 \section{Notation and Formulae}

In this section a very brief sketch will be given of the
cosmological model and the notation.  For details see Humphreys et
al \cite{DR}. In observational coordinates the spherically
symmetric model takes the form \cite{RM},
 \begin{equation}
 ds^{2} = - A(w,y)^{2} dw^{2} + 2 A(w,y) B(w,y) dw dy + C(w,y)^{2}
  d \Omega^{2} \label{SP}
 \end{equation}
where \{$w = $ const \} label the past light cones on the
observer's world line \{$ y = 0$ \}, $y$ is a comoving radial
distance coordinate down the null geodesic \{$w = $ constant,
$(\theta, \phi) = $ const \} and $ d \Omega^{2} = d \theta^{2} +
\sin^{2} \phi d \phi^{2}$.  The dust velocity is given by
$u^{i} = A^{-1} \delta^{i}_{w}$. The FLRW models are given by
choosing $A = B$.  In $(3 + 1)$ coordinates the model is given by,
 \begin{equation}
ds^{2} = - dt^{2} + \left[\frac{\partial R(r,t)}{\partial
r}\right]^{2} \frac{dr^{2}}{1 - kf(r)^{2}} + R(r,t)^{2} d
\Omega^{2}  \label{OC}
 \end{equation}
where $f$ is arbitrary and relates to the total energy of the
system.  The geometry of the $\{t = $ const $\}$ hyper-surfaces is
parabolic if $ k = 0$, hyperbolic for $k = - 1$ and elliptic for $k
= 1$.  The dust velocity is, $ u^{i} = \delta^{i}_{t}$.

From the two metrics the subtle relations between the two systems
can be demonstrated.  On a fixed light cone $\{w = w_{0}\}$ the
observer area distance (angular diameter distance) $D$ in terms of
redshifts is given by, \\
\begin{center}
 $ D(y(z))  =  \left\{ \begin{array}{c} C(w_{0}, y(z)) \\
                 R\left(r(z), T(r(z))\right)
 \end{array} \right. $
 \\
\end{center}
\noindent where $t = T(r)$ is the equation of the past light cone of 
the
observer.  In general the function $T$ cannot be determined as an
exact function and so numerical methods have to be employed
\cite{RI} and thus to determine $R$ we need to know both $r$ and
$t$ separately.  Mustapha et al \cite{MU} circumvent this by
restricting attention to a single light cone, which cuts out the
dynamics,  and they use time $t$ which is not an observable. These
are not appropriate here.

In the metric (\ref{SP}) the hyper-surfaces \{$w = $ const\} are
null but that is not sufficient to make them the past null cones of
the observer.  This requires central conditions \cite{ST} which
develops earlier results of Temple \cite{TE}.  As $y
\rightarrow 0$ the following asymptotic behaviour is required,
 \begin{eqnarray}
 A(w,y) & = & A(w,0) + O(y) \\
 B(w,y) & = & B(w,0) + O(y) \\
 C(w,y) & = & B(w,0)y + O(y^{2})
 \end{eqnarray}
 with $A(w,0) \neq 0$ and $ B(w,0) \neq 0$.  These conditions imply 
 that near
 enough to the observer the space-time appears Minkowskian.  This
 imposes restrictions on the asymptotic behaviour of the number count 
 formula as
 $ y \rightarrow 0$.  In observational coordinates the number count 
 formula is
 given by \cite{ST},
  \begin{equation}
  N(y) = 4 \pi \int^{y}_{0} n(w_{0}, x) B(w_{0}, x) C(w_{0}, x)^{2} d 
x
   \end{equation}
   where $n(w_{0}, y)$ is the number density of sources.  The dust 
   density is
   \begin{equation}
   \rho(w_{0}, y) = n(w_{0}, y) m
    \end{equation}
    where $m$ is the average mass of the sources (galaxies).  We 
ignore 
    selection
    and evolution effects although they could be incorporated without 
    much
    difficulty if we knew their functional form.  Also we assume no 
bias.  
    Dark matter
    could be included via $m$,  if unbiased.  From the energy 
    conservation
    equation, the central conditions and the formulae for $N$ 
    and $\rho$ it follows
    that as $y \rightarrow 0$,
     \begin{equation}
\rho = 4\pi \rho_{0}\frac{B(w_{0}, 0)^{3}}{B(w,0)^{3}} + O(y)
     \end{equation}
and
 \begin{equation}
 N(y) = \left( \frac{4 \pi \rho_{0}}{3m}\right) y^{3} + O(y^{4})
 \end{equation} where $\rho_{0} = mn(w_{0}, 0)$

 These strictly mathematical results mean that the number count 
 formula cannot be
 fractal for $y$ near zero.
  
  \section{Sketch to Illustrate the Relations Between the Variables}

From the two formulae for the dust velocity we obtain,
 \begin{eqnarray}
 A & = & \frac{\partial t}{\partial w} \label{A} \\
 B & = & - \frac{\partial t}{\partial y}
  \end{eqnarray}
  and these can be used to write the field equation for $A$ in the 
form 
  \cite{RM},
  \begin{equation}
  A = \frac{\dot{C}}{\left[(F^{2} - 1) + \left( 
\frac{2mN_{*}}{C}\right)
  \right]^{1/2}} \label{C}
   \end{equation}
   where $\dot{C} = \partial C/\partial w $, 
   $N^{\prime}_{*} = F N^{\prime} $, $F$
   is an arbitrary function of integration and
   a prime denotes the partial derivative with respect to $y$.  We 
can use
   (\ref{A}) to write (\ref{C}) in the form,
    \begin{equation}
 dt = \frac{ d C}{\left[(F^{2} - 1) + \left( \frac{2mN_{*}}{C}\right)
  \right]^{1/2}} \label{DC}
     \end{equation}
     along the fluid flow field.  Along this flow the well known 
Tolman 
     solution
     is \cite{BN} \cite{TO}
     \begin{equation}
  dt = \frac{dR}{\left[- kf + \frac{2M}{R} \right]^{1/2}}
     \end{equation}
     which led us to identify,
  \begin{eqnarray}
  F(y)^{2} & = & 1 - k f(y)^{2} \\
  M(y) & = & mN_{*}(y) \\
       & = & m \int^{y}_{0} (1 - k f(y)^{2})^{\frac{1}{2}} 
       \frac{dN}{dx}dx  \label{M}
  \end{eqnarray}
Clearly for $k = 0$, $M(y) = m N(y)$.

To obtain further equations, e.g. for $(1 - k f(y)^{2})$, in terms
of observational data $D(z)$ and $N(z)$ requires further use of the
field equations, given by Maartens et al \cite{RM}.  This results
in,
 \begin{eqnarray}
 \sqrt{1 - k f(z)^{2}} & = & \frac{1 + z}{2D(z)} \int^{z}_{0} 
\frac{1}{Q} 
 \left\{
D^{\prime}(x) + \left[\frac{D(x)Q(x)^{2}}{(1 +
x)^{2}}\right]^{\prime}  \right\} dx \\
 Q(z) & = & 1 - m\int^{z}_{0} \frac{(1 + x) N^{\prime}(x)}{D(x)} dx.
 \end{eqnarray}
 The full set of equations is given by Humphreys et al \cite{DR}.  
 The equation
 (\ref{M}) can be inverted to give,
  \begin{equation}
  N(y) = \frac{1}{m} \int^{y}_{0} M^{\prime}(x) 
  [1 - k f(x)^{2}]^{-\frac{1}{2}} dx
   \end{equation}
   \section{An Application of the Formulae}

To apply the formulae we only need to consider one past light cone
so we can use the area distance $D$ as our distance measure.  We
will also limit ourselves to the $k = 0$ case which is particularly
simple because only one of the functions $N$ and $D$ is arbitrary.
The other more complicated cases which cannot be handled completely
analytically are considered by Humphreys et al \cite{DR}.

The dependence of $N$ and $D$ is expressed explicitly by the
integral equation \cite{DR},
 \begin{equation}
1 + z = \left(1 - \sqrt{\frac{2mN}{D}}\right)^{-1}\exp\left[- \int
\frac{m}{D}\frac{dN}{dD}\left(1 - \sqrt{\frac{2mN}{D}} \right)^{-1}
dD \right]  \label{shift}
 \end{equation}
 Once $N$ is known in terms of $D$, through for instance a fractal
 formula, this gives a relatively simple expression for $z$ in terms
 of $D$.  Unfortunately the central conditions do not permit such a
 simple procedure since the form of the function $N$ has to change.
 A simple ansatz \cite{DR} which combines the limiting behaviour
 with a fractal number count formula outside the immediate vicinity of
 $D = 0$, is
 \begin{equation}
  N(D) = \left\{ \begin{array}{ll}
        \frac{4\pi \rho_{0}}{3m}D^{3} & \mbox{for $ D 
        \leq D_{I}$} \\
        \frac{4 \pi \rho_{0}}{3m} D_{I} \left( 
        \frac{D}{D_{I}}\right)^{3}
        & \mbox{$ D \leq D \leq D_{h}$}
     \end{array}\right. \label{mod}
  \end{equation}
where $D_{h}$ is the distance at which transition to FLRW geometry
occurs.  We will assume that in the core $(D \leq D_{I})$ and in
the fractal region $(D_{I} < D \leq D_{h})$ the space-time is
parabolic, i.e., $k = 0$.  Then the Hubble constant is given by
\cite{RM},
 \begin{equation}
H_{0} = \sqrt{\frac{8}{3}\pi \rho_{0}}
 \end{equation}
 which puts a precise constraint on the central density.  It can be 
 shown \cite{DR}
 that matching the number count $N(y)$ at $D_{h}$ relates 
 the four parameters
 $H_{0}, D_{I}, D_{h},$ and $\nu$.  From observational data one 
 could estimate
$H_{0}, D_{h}$ and $\nu$ which would then fix the minimum fractal
scale $D_{I}$.  Unfortunately the critical observations are
controversial.

For small $D$ an explicit form of the behaviour of $z$ with respect
to $D$ is given by \cite{DR},
 \begin{equation}
\frac{dz}{dD} \approx \frac{1}{2}H_{0} \left[(\nu - 1)\left( 
\frac{D_{I}}{D}
\right)^{(3 - \nu)/2} - \; \nu H_{0} D_{I} \left(\frac{D_{I}}{D}
\right)^{(2 - \nu)} \right],
\end{equation}
which shows that after initial behaviour which is linear by
construction to $D_{I}$ the $z(D)$ graph curves upwards
contradicting the linear Hubble Law for scales less than 100 Mpc.
This rules out the parabolic models.

It is interesting that for any rational fractal index $\nu$
(\ref{shift}) can be integrated to give an explicit expression for
$z$ in terms of $D$.  Note that for the simple model of the
transition from the core to the fractal region given in (\ref{mod})
the integration is particularly simple but it can become very
complicated for more sophisticated transition formulae. In the
non-parabolic models it can be shown \cite{DR} that the fractal
number count forces a very low density for the universe.
\section{Conclusion}

The talk outlined some of the ideas and methods employed by
Humphreys et al \cite{DR}. Only some aspects have been discussed to
show the mathematical ideas involved.  The comprehensive treatment
\cite{DR} concludes that exact models using fractal counts models
out to 150 Mpc either contradict the well established linear Hubble
law out to 150 Mpc or they yield a very low density universe. Both
conclusions conflict with observations.  There are various
modifications that could be made to the model but the most
compelling is that the luminous matter may not trace the actual
distribution of density.  In this connection a recent paper by
Labini and Durer \cite{DU} argues for a fractal baryonic matter
distribution with non fractal dark matter.  The approach developed
here gives an alternative method by which bias could be detected or
measured.

 \section*{Acknowledgements}
 This research is supported by the University of Portsmouth.  The 
author
 thanks Dr Roy Maartens for valuable comments on a draft of the paper.

 \end{document}